\documentclass{llncs}
\usepackage[utf8]{inputenc}
\pdfoutput=1

\usepackage{subfigure}
\usepackage{graphicx}

\usepackage{textcomp} 
\usepackage{amsmath,amssymb}
\usepackage{textgreek}

\usepackage[bookmarks, colorlinks=true, linktoc=page, pdftex, linkcolor=blue, citecolor=blue, urlcolor=blue]{hyperref}

\usepackage{texnames}
\usepackage{lineno}
\usepackage{authblk}

\begin{document}

\title{The CMS High-Granularity Calorimeter for Operation at the High-Luminosity LHC}
\author{Florian Pitters$^{1,2}$ \\ \small{On behalf of the CMS collaboration}}
\institute{CERN, Geneva, Switzerland \and Vienna University of Technology, Vienna, Austria}

\maketitle

 
\begin{abstract}
The High Luminosity LHC (HL-LHC) will integrate 10 times more luminosity than the LHC, posing significant challenges for radiation tolerance and event pileup on detectors, especially for forward calorimetry, and hallmarks the issue for future colliders. As part of its HL-LHC upgrade program, the CMS collaboration is designing a High Granularity Calorimeter to replace the existing endcap calorimeters. It features unprecedented transverse and longitudinal segmentation for both electromagnetic (ECAL) and hadronic (HCAL) compartments. This will facilitate particle-flow calorimetry, where the fine structure of showers can be measured and used to enhance pileup rejection and particle identification, whilst still achieving good energy resolution. The ECAL and a large fraction of HCAL will be based on hexagonal silicon sensors of 0.5 to 1\,cm$^2$ cell size, with the remainder of the HCAL based on highly-segmented scintillators with SiPM readout. The intrinsic high-precision timing capabilities of the silicon sensors will add an extra dimension to event reconstruction, especially in terms of pileup rejection. An overview of the HGCAL project is presented, covering motivation, engineering design, readout and trigger concepts, and expected performance.
\end{abstract}

\begin{keywords}
CMS; HGCAL; calorimeters; silicon pad detectors; high granularity; particle flow.
\end{keywords}



\section{Introduction}
Starting from 2026 onwards, the HL-LHCs instantaneous luminosity will be increased by a factor 5 to 7 compared to LHC and will result in up to 200 collisions per bunch crossing. In this mode, LHC will run for 10 years and deliver an integrated luminosity of about 3000\,fb$^{-1}$. The current CMS detector was designed for operation at 25 collisions per bunch crossing and up to 500\,fb$^{-1}$~\cite{CMS2008}. 

\begin{figure}[t]
    \subfigure[]{
        \includegraphics[width=0.49\textwidth]{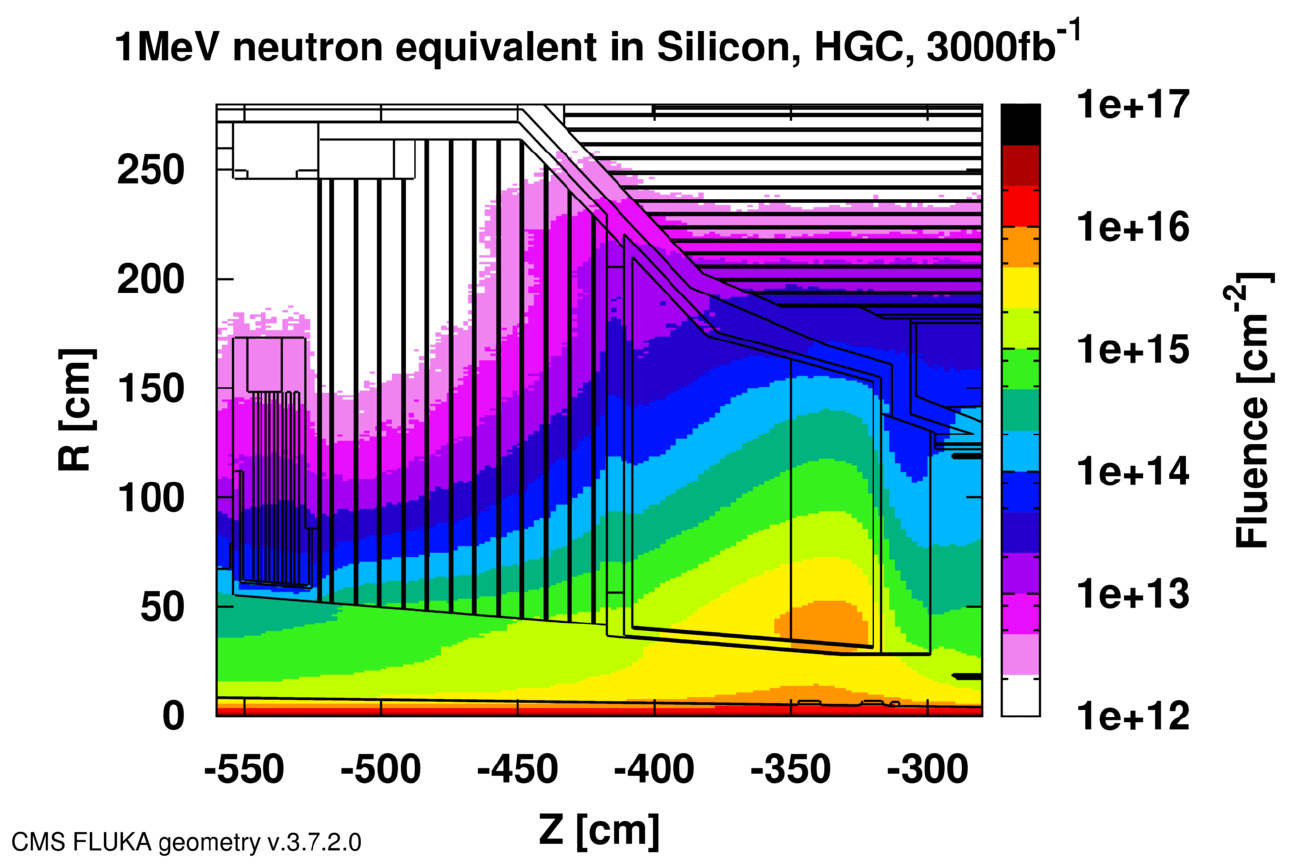}
        \label{fig:fluence}}
    \hfill
    \subfigure[]{
        \includegraphics[width=0.493\textwidth]{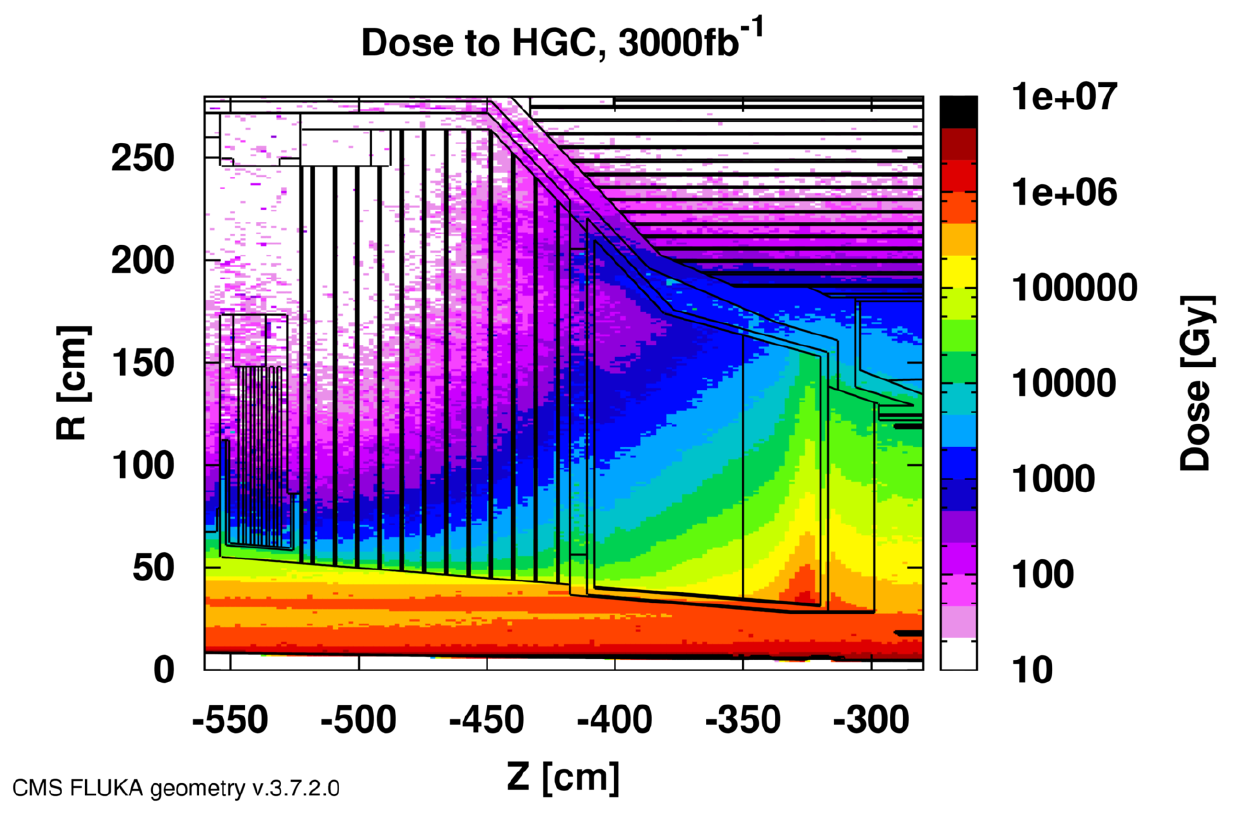}
        \label{fig:dose}}
    \caption{The expected integrated hadron fluences for the endcap expressed in 1\,MeV neutron equivalent per cm$^2$ are shown in (a) and the total integrated dose in (b). The flux and dose are varying with R and Z, allowing for different technology choices depending on the exact location. The electromagnetic part of HGCAL will use silicon as active medium while the hadronic part will use silicon in the innermost regions and scintillating tiles with SiPM readout for the outer parts. Figures first printed in Ref.~\cite{CMS2015}. Published with permission by CERN.}
    \label{fig:doses}
\end{figure}

To cope with the new environment and retain a good physics performance up to 3000\,fb$^{-1}$, several upgrades to the CMS subdetectors are planned~\cite{CMS2015}. The endcap calorimeters are among the subdetectors that will be most exposed to high radiation levels. Fig.~\ref{fig:doses} shows the expected total dose and hadron fluences as a function of R and Z. In the innermost regions, the detector has to withstand 10$^{16}$\,neq/cm$^2$ and 150\,MRad. Under these conditions, the current endcap calorimeters would degrade very quickly in performance~\cite{CMS2015}. Therefore, they will be completely replaced by a silicon and scintillator based highly granular sampling calorimeter called HGCAL (High Granularity Calorimeter). 



\section{Detector Design}
The CMS HGCAL consist of an electromagnetic part called EE and two hadronic parts called FH \& BH.\footnote{EE stands for ``Endcap Electromagnetic'' calorimeter, FH for ``Front Hadronic'' calorimeter and BH for ``Back Hadronic'' calorimeter.} The electromagnetic part will be 25\,X$_0$ deep and will consists of 28 layers of silicon pad sensors as active elements with lead in a stainless steel envelope as absorber. The two hadronic parts are in total 8.5\,\textlambda$_I$~deep with 24 layers and steel absorbers. As active elements, silicon will be used in the high $|\eta|$ regions and scintillating tiles with SiPM readout in the lower $|\eta|$ regions. The full system will be maintained at -30$^\circ$C using evaporative CO$_2$ cooling to limit the leakage current of the silicon sensors. 

With silicon pads and scintillating tiles, high granularity in transverse and longitudinal direction will be maintained throughout the calorimeter and will allow for particle flow analysis. High precision time measurement with better than 50\,ps resolution on a cell level is aspired for vertex reconstruction and pile-up rejection.

\subsection{Active Elements}
\begin{figure}[t]
    \subfigure[]{
        \includegraphics[width=0.47\textwidth]{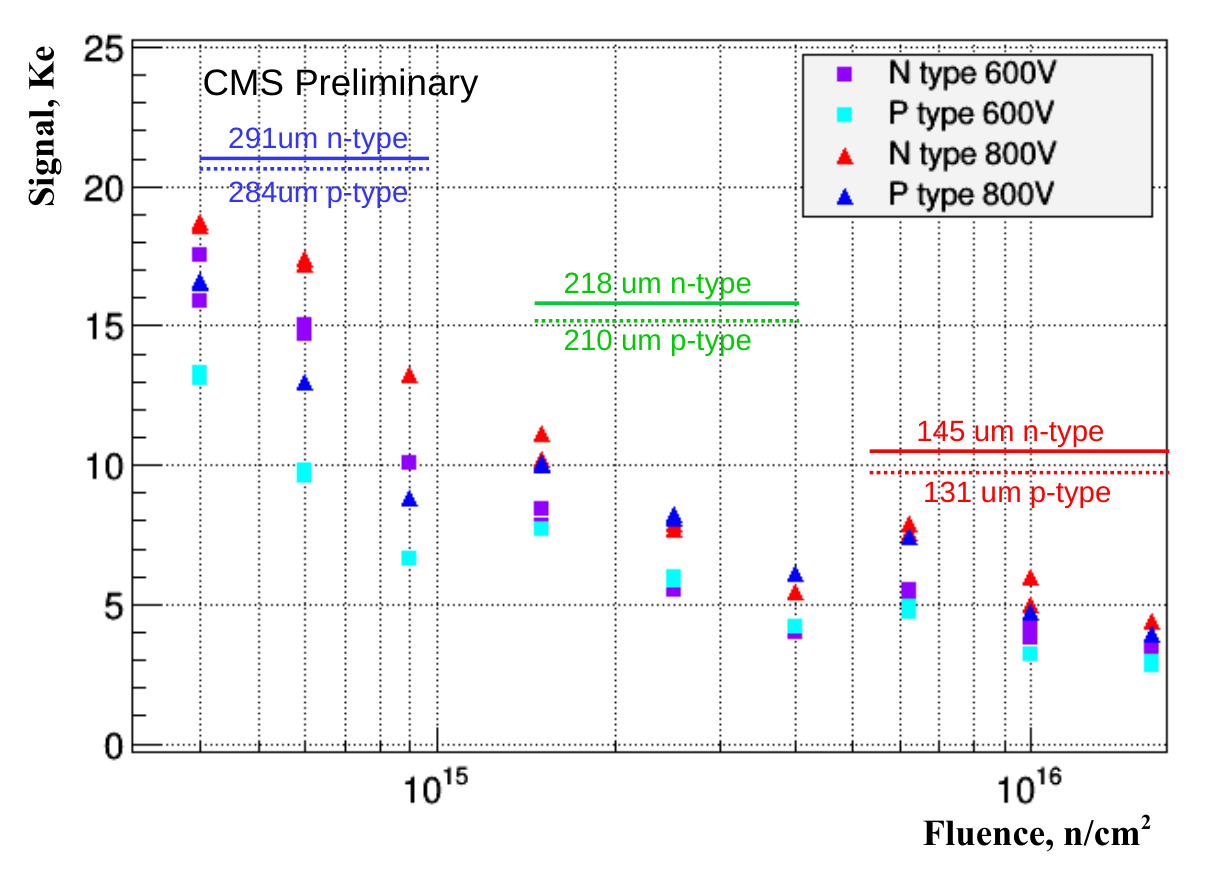}
        \label{fig:loss}}
    \hfill
    \subfigure[]{
        \includegraphics[width=0.53\textwidth]{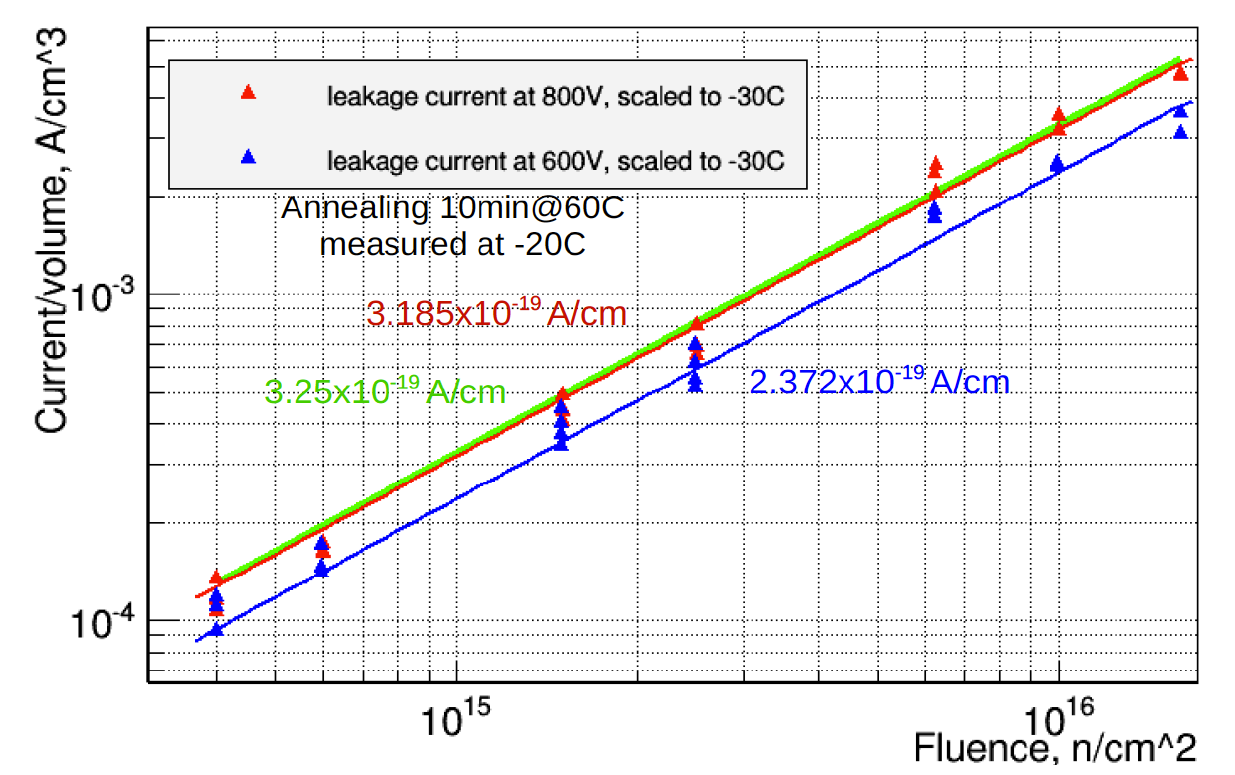}
        \label{fig:leakage}}
    \caption{The mean signal in silicon diodes for different neutron fluences can be seen in (a). Thinner sensors and operation at higher voltages mitigate the signal loss. The scaling of leakage current with active detector volume and neutron fluence is shown in (b). The noise contribution scales with the square root of the leakage current. Figures first printed in Ref.~\cite{CMS2015}. Published with permission by CERN.}
    \label{fig:signal}
\end{figure}

One of the most relevant quantity for the detector performance is the signal-to-noise ratio. For silicon, it has been shown that the signal loss due to irradiation is decreased in thinner sensors and when operating at increased bias voltages~\cite{CMS2015,Curras2016}. The increased noise contribution from the leakage current can be mitigated by cooling. Both aspects are displayed in Fig.~\ref{fig:signal}. Additionally, the intrinsic time resolution of silicon has been shown to be below 15\,ps for signals above 20 MIPs~\cite{Curras2016}.

In total, the system will consist of roughly 600\,m$^2$ of silicon. The use of 6 or 8 inch wafers with hexagonal geometry is foreseen to reduce costs. The active thickness will be adapted to the expected radiation dose and will vary between 120, 200 and 300\,\textmu m.\footnote{Whether the active thickness is best reached via deep diffusion, physical thinning or an epitaxial layer is currently under study.} The cell capacitance should be around 50\,pF for all sensor thicknesses and therefore thinner sensors will be equipped with smaller cells. A granularity of 0.5\,cm$^2$ for the 120\,\textmu m and 1~cm$^2$ for 200 and 300\,\textmu m thick sensors will be used. One of the key aspects of these sensors is the high-voltage sustainability to mitigate radiation damage. The goal is a breakdown voltage above 1\,kV. It is also foreseen to use a few cells with smaller area than the regular ones on each sensor. The smaller area at unchanged thickness will reduce the noise contributions from capacitance and leakage current in these cells, so that they should still be sensitive to single MIPs after 3000\,fb$^{-1}$.


At larger distances to the interaction point radiation levels are lower and plastic scintillating tiles with SiPM readout will be used, analogous to the CALICE AHCAL~\cite{CALICE}. The exact intersection between scintillator and silicon regions as well as the tile granularity will be evaluated in the coming months.

\subsection{Modules, Absorbers and Mechanical Integration}
Silicon modules start with a metallic baseplate (CuW in EE and Cu in FH/BH), which acts as an absorber and mechanical support, that has a polyimide gold-plated foil glued to it. The silicon sensor is then glued onto that foil. The readout PCB hosting the front-end ASICs is in turn glued onto the sensor and wirebonds reaching through holes in the PCB connect to the sensor contact pads. The design of the active scintillator modules is currently being developed. 

Modules will be mounted on cooling plates together with the front-end electronics to make up cassettes. The absorber structure that hosts the cassettes will be made in full disks to guarantee an optimal physics performance. In the EE case, self-supporting double sided cassettes are used while in the FH and BH case, the mixed cassettes will be directly mounted on the steel absorber.

\subsection{Readout Electronics}
The driving requirements for the front-end readout ASIC are a large dynamic range of 0.4\,fC to 10\,pC (15 bits), a noise level below 2000 electrons, timing information with below 50~ps accuracy and radiation hardness up to 150\,MRad. The goal is to keep within a power budget of around 10\,mW/channel for the analog part. To meet these requirements, a chip based on OMEGA's ROC family~\cite{OMEGA} is being developed. The baseline option includes two traditionl gain stages and a time-over-threshold stage, as well as a time-of-arrival path with 50\,ps binning. The ASIC will be fabricated in TSMC 130\,nm CMOS technology which has been qualified up to 400\,MRad~\cite{CMS2015}.

Information from HGCAL will also be used for the L1 trigger decision. A subset of the data is sent to a concentrator chip and, after clustering, combined with the track trigger. The trigger latency of 12.5\,\textmu s drives the requirement for large buffer sizes in the readout chip.

A first version of the readout chip will be submitted in the summer of 2017.


\section{Expected Performance}
The choice of lead as absorber with a small Moliere radius and a large ratio of interaction length to radiation length allows for a compact calorimeter with excellent particle separation capabilities. The narrow showers together with the high granularity and excellent time resolution will allow for a pile-up suppression in the first few layers of EE. The instrinsic energy resolution of the EE part for incident electrons is expected to have a stochastic term below 25\%/$\sqrt{\text{GeV}}$ and a constant term below 1\%~\cite{CMS2015}. These values are sufficient as the energy resolution will be dominated by the confusion term in the particle flow algorithm rather than the intrinsic resolution of the calorimeter. Optimisation of these algorithms to the physics environment and detector design is currently ongoing.


\section{Outlook}
The CMS collaboration is making good progress towards the construction of a new generation of imaging calorimeter. The basic design has been validated in testbeam and design optimisation is ongoing. The technical design report is expected to be released by the end of 2017.



\bibliographystyle{splncs03}

\end{document}